\documentclass[aps,prl,twocolumn,superscriptaddress]{revtex4}
\usepackage{amsmath}
\usepackage{amssymb}
\usepackage{graphicx}

\begin{document}

\title{Analytical canonical partition function of a quasi-one dimensional
system of hard disks}
\author{V.M.~Pergamenshchik}
\email{victorpergam@yahoo.com}
\affiliation{Institute of Physics, prospekt Nauky, 46, Kyiv 03039, Ukraine}
\date{\today }

\begin{abstract}
The exact canonical partition function of a hard disk system in a narrow
quasi-one dimensional pore of given length and width is derived analytically
in the thermodynamic limit. As a result the many body problem is reduced to
solving single transcendental equation. The pressures along and across the
pore, distributions of contact distances along the pore and disks'
transverse coordinates are found analytically and presented in the whole
density range for three different pore widths. The transition from the
solidlike zigzag to liquidlike state is found to be quite sharp in the
density scale but shows no genuine singularity. This transition is
quantitatively described by the distribution of zigzag's windows through
which disks exchange their positions across the pore. The windowlike defects
vanish only in the densely packed zigzag which is in line with a continuous
Kosterlitz-Thouless transition.\newline

*email: victorpergam@yahoo.com

This paper is published in \textbf{J. Chem. Phys. 153, 144111 (2020)};
https://doi.org/10.1063/5.0025645 
\end{abstract}

\maketitle

\section{Introduction}

Over more than a century the idea to model molecules as hard spheres has
been widely used in the theory of liquids \cite{Hansen,Yukhnovski,HS}. In
spite of apparent simplicity, the behavior of hard sphere systems is so
complex mathematically that no exact analytical result has been obtained in
3 and even in 2 dimensions (2D). Under these circumstances, the numerical
Monte Carlo and molecular dynamics approaches have become the main tools in
the study of 3D hard sphere and 2D hard disk (HD) systems. The numerical
results however are restricted to systems of a finite number of particles
whereas such effects as, for instance, phase transitions, are related to
systems in the thermodynamic limit when the number of particles $N$ is
infinite. As this limit can be studied only theoretically, analytical
results are of great importance. The study of hard sphere systems is not
restricted to their statistical equilibrium. Nowadays this system is also
considered as a useful model glass former, and much efforts in this area
have been devoted to analytical solutions in high unphysical and even
infinite dimensions in a hope to get an insight into hard sphere systems in
physical dimensions 2 and 3 (for instance, recent paper \cite{Hicks} and
review \cite{Glass}). But to attack the real dimensions directly is very
difficult. The first exact analytical result was obtained in 1936 by Tonks
for the purely 1D system of HDs. This system is much simpler than any 2D
system, nevertheless Tonks' solution has become the analytical platform for
further expansion into the world of 2D HD systems via moving to certain
quasi-1D models. Barker was the first to point to the general possibility
that the 1D case is amenable to a solvable generalization to quasi-1D case
of HDs in narrow pores \cite{Barker}. The simplest quasi-1D system (from now
on just q1D) is such that each disk can touch no more neighbors than one
from both sides (the so-called single-file system); the width of such q1D
pore must be below $\sqrt{3}/2+1$ times HD diameter. The analytical theory
of HDs in q1D pore was presented by Wojciechovski et al \cite{Wojc} and ten
years later was further developed by Kofke and Post \cite{Kofke}. Kofke and
Post have elegantly shown that the problem can be reduced to solving certain
integral equation. In general however the integral equation of this, now
known as the transfer matrix method, cannot be solved analytically. A
density expansion \cite{Kamen} and simplified model \cite{Varga} to
approximately solve this equation analytically have been proposed. The
peculiarity of this method is that it is essentially related to the
pressure-based ($N,P,T)$ ensemble which does not directly predict pressure
as a function of the system's width $D$ and length $L$. In pursuit of an
analytical result, the virial expansion for a q1D HD system has also been
developed up to the fourth term \cite{Mon}. In this paper I present exact
analytical derivation of the canonical ($N,L,D,T)$ partition function (PF)
in the thermodynamic limit. As a result, finding the thermodynamic
properties of a q1D HD system for given $L$ and $D$ is reduced to solving
single transcendental equation which can be easily done numerically. The PF,
pressure along and across the pore, distribution of the contact distances
between neighboring HDs along the pore, and distribution of HD centers
across the pore are found analytically; the constant to be found numerically
fully specifies the PF and these distribution functions. The longitudinal
and transverse pressures and the above distributions are presented for three
pore widths in the total range of linear density $\rho =N/L$. The disks'
arrangement for different densities $N/L$ is discussed.

The developed analytical theory enables for a deeper insight into the
transition from solidlike to liquidlike state in a q1D HD system. The system
shows a sharp crossover, but the thermodynamics does not show any genuine
discontinuity. We found that this is similar to the melting in a continuous
Kosterlitz-Thouless-type transition. A solid-to-fluid transition is a global
phenomenon attributed to the entire body, but in a 2D crystals it starts
from local emergence of bounded defect pairs \cite{KT}. The densely packed
state of a q1D system is the zigzag array where all disks are caged and
cannot move across the pore. To gain entropy the system searches for
uncaging. Though uncaging cannot occur in the entire system, it can occur
locally where pair of disks tries to exchange their positions across the
pore. Usually, the density of defects is determined by their core energy via
Boltzmann's factor, but in our case it is irrelevant as possible defects
have a purely entropic origin. The distribution of contact distances along
the pore gives the rate of such entropic defects. As density decreases, it
predicts an emergence of progressively larger number of windowlike defects
in the zigzag arrangement. These windows are of the size of disk diameter so
that through them pairs of disks can exchange their position across the
pore. The number of these defects vanishes only in the densely packed state
which is in line with a continuous Kosterlitz-Thouless-type transition. The
similarity has been strongly supported by the results on the correlation
decay in a q1D HD system recently obtained by molecular dynamics simulation\
in \cite{A i T,WE}. Note that defects of the zigzag arrangement, which are
similar to our windowlike defects, in connection with their role in the disk
motion across the pore have been discussed earlier in \cite{Hicks,Robinson}.

\section{Partition function}

Consider a pore of the width $D$ and length $L$ filled with $N$ HDs of
diameter $d=1,$ Fig.1. We assume the thermodynamic limit $N\rightarrow
\infty ,L\rightarrow \infty $ while $N/L=const$; the terms which vanish in
this limit (e.g., the end effects) will be omitted. All lengths will be
measured in HD diameters. The width parameter $\Delta =(D-d)/d$, that gives
the actual pore width attainable to HD centers, in the quasi 1D case ranges
from $0$ in the 1D case to the maximum $\sqrt{3}/2\approx 0.866.$ The $i$-th
disk has two coordinates, $x_{i}$ along and $y_{i}$ across the pore; $y$
varies in the range $-\Delta /2\leq y\leq \Delta /2;$ the pore volume is $%
LD. $ The vertical center-to-center distance between two neighbors, $\delta
y_{i}=y_{i+1}-y_{i},$ determines the contact distance $\sigma _{i}$ between
them along the pore: 
\begin{eqnarray}
\sigma _{i} &=&\min \left\vert x_{i+1}(y_{i+1})-x_{i}(y_{i})\right\vert , 
\notag \\
\sigma _{i} &=&\sqrt{d^{2}-\delta y_{i}^{2}},  \label{sigma} \\
\sigma _{m} &=&\sqrt{d^{2}-\Delta ^{2}}\leq \sigma \leq d=1.  \notag
\end{eqnarray}

The minimum possible $\sigma ,\sigma _{m},$ obtains for $\delta y=\pm \Delta 
$ when the two disks are in contact with the opposite walls. Thus, each set
of coordinates $\{y\}=$ $y_{1,}y_{2},...,y_{N\text{ \ }}$determines the
correspondent densely packed state of the total length $L^{\prime
}\{y\}=\sum_{i=1}^{N-1}\sigma _{i}(\delta y_{i}),$ Fig. 1, which we call
condensate and which plays the central role in our theory. The minimum
condensate length is $\sigma _{m}N,$ the maximum length can be as large as $%
Nd$, but it cannot exceed $L:$ $N\sigma _{m}<L^{\prime }\leq L_{\max
}^{\prime }$ where $L_{\max }^{\prime }=\min (Nd,L).$

\begin{figure}[tbp]
\includegraphics[height=10cm]{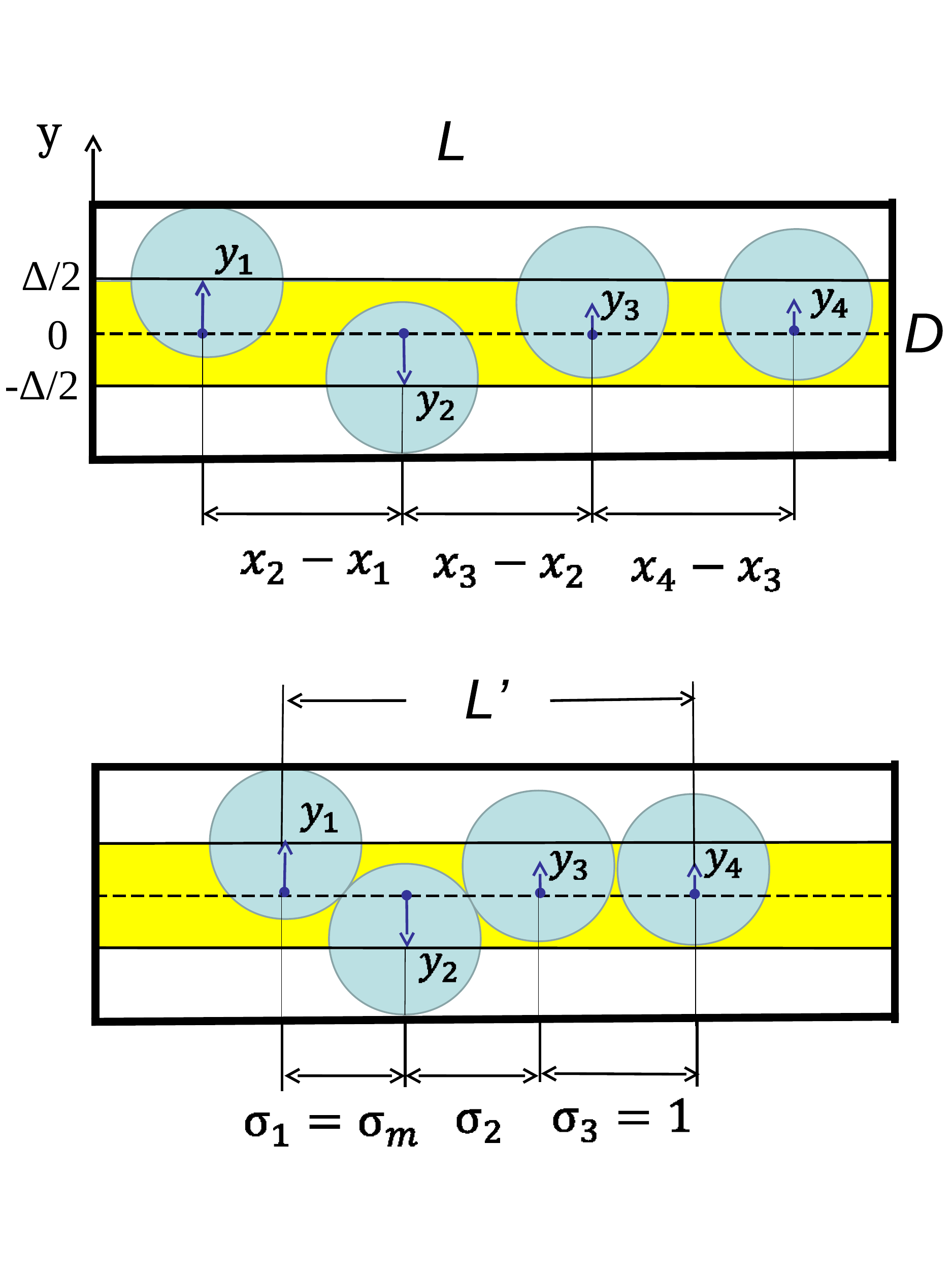}
\caption{{}Four HDs in the $L\times D$ pore and the condensate (below)
correspondent to their vertical coordinates $\{y\}.$ The inner fraction of
thickness $\Delta $ is the volume accessible to disks' centers. The end
effect in $L^{\prime }$ (one diameter $d$ compared to $L^{\prime })$ is
neglected.}
\label{Fig1}
\end{figure}

In this paper we will systematically use singular functions and their
analytical representations as this substantially simplifies both the
integration domains and calculation of the manyfold integrals. We will use
the notation $D^{N-1}x=dx_{1}...dx_{N-1}$ for the product measure. Omitting
unimportant factors, the exact configurational canonical $(N,L,D)$ PF of the
q1D\ HD system has the following form: 
\begin{eqnarray}
Z &=&\int\limits_{-\Delta /2}^{\Delta /2}D^{N}y\times \theta \left( L_{\max
}^{\prime }-\sum_{i=1}^{N-1}\sigma _{i}\right)  \label{zz} \\
&&\times \theta \left( \sum_{i=1}^{N-1}\sigma _{i}-(N-1)\sigma _{m}\right)
\int_{X}D^{N}x,  \notag
\end{eqnarray}%
where $\theta $ is the step-function: $\theta (x)=1$ for $x\geq 0$ and $%
\theta (x)=0$ otherwise. The $x$ integration domain $X$ was first formulated
by Tonks \cite{Tonks} for the 1D case and much later by Wojciechovski et al 
\cite{Wojc} for a q1D system:%
\begin{equation}
\int_{X}D^{N}x=\int\limits_{d/2}^{x_{2}-\sigma
_{1}}dx_{1}\int\limits_{\sigma _{1}+d/2}^{x_{3}-\sigma
_{2}}dx_{2}...\int\limits_{\sum_{i=1}^{N-2}\sigma _{i}+d/2}^{L-d/2}dx_{N}
\label{Dx1}
\end{equation}%
For all $y$ coordinates fixed, $X$ ensures that under the single-file
condition the disks do not intersect: disk $i$ can move between two next
neighbors $i-1$ and $i+1,$ its minimum distance to disk $i-1$ is $\sigma
_{i-1}(\delta y_{i-1})$ and that to disk $i+1$ is $\sigma _{i}(\delta
y_{i}), $ and so on; the minimum distance between disks $1$ and $N$ and the
correspondent walls is $d/2.$ We start with resorting to the alternative
form of integral (\ref{Dx1}) in which the integration domain is fixed by a
theta function. To this end we change from the variables $%
x_{1},x_{2},...,x_{N}$ to variables $x_{1},\delta x_{1}\delta
x_{2},...\delta x_{N-1}$ where $\delta x_{i}=x_{i+1}-x_{i};$ the Jacobian of
this well-known change of variables is 1. Then the $x$ integral takes the
following form:

\begin{eqnarray}
\int_{X}D^{N}x &=&\int\limits_{0}^{L-d-\sum_{i=1}^{N-1}\sigma _{i}}\frac{%
dx_{1}}{N!}  \label{Dx2} \\
&&\times \int\limits_{0}^{L-d-\sum_{i=1}^{N-1}\sigma _{i}}D^{N-1}\delta
x\theta \left( L-\sum_{i=1}^{N-1}\delta x_{i}\right) .  \notag
\end{eqnarray}%
The theta fuction $\theta \left( L-\sum \delta x_{i}\right) $ restricts the $%
x$ integration to those $x$ for which the condensate's length does not
exceed the total length $L;$ as each $\delta x(\delta y)$ is restricted by
its lower boundary $\sigma $($\delta y)$ the disks cannot overlap. As the
integrarion limits in all $\delta x$ are the same, the factor $1/N!$ is
needed to exclude permutations of the $x$ coordinates which were precluded
in the original integral (\ref{Dx1}) by the form of the integration domain.
The analytical form of the theta function is%
\begin{equation}
\theta (x)=\frac{1}{2\pi }\int\limits_{-\infty }^{\infty }\frac{d\alpha }{%
i\alpha }e^{i\alpha x},  \label{Tet}
\end{equation}%
where the integration path circumvents the point $\alpha =0$ from below. We
make use of this formula in (\ref{Dx2}) and peform the $\delta x$
integration to get%
\begin{eqnarray}
\int_{X}D^{N}x &=&\int\limits_{0}^{L-d-\sum_{i=1}^{N-1}\sigma _{i}}\frac{%
dx_{1}}{N!}\int\limits_{-\infty }^{\infty }\frac{d\alpha }{2\pi i\alpha }%
e^{i\alpha L}  \label{Dx3} \\
&&\times \left[ \frac{\exp i\alpha \left( L-d-\sum_{i=1}^{N-1}\sigma
_{i}\right) -1}{i\alpha }\right] ^{N-1}.  \notag
\end{eqnarray}%
The $\alpha $ integrand has a pole at $\alpha =0$ which is of the first
order as the expression in the square brackets is regular at this point.
Taking the residue and performing the $x_{1}$ integration one finally
obtains:%
\begin{equation}
\int_{X}D^{N}x=\frac{1}{N!}\left( L-d-\sum_{i=1}^{N-1}\sigma _{i}\right)
^{N}.  \label{Dx0}
\end{equation}%
In the 1D case, all $\sigma $'s are equal to $d$\ and this expression is the
PF obtained by Tonks:%
\begin{equation}
Z_{1D}=\frac{1}{N!}\left( L-Nd\right) ^{N}.  \label{Z1D}
\end{equation}%
Let us proceed with the PF (\ref{zz}). Substituting (\ref{Dx0}) and changing
from the variables $y_{1,}y_{2,...,}y_{N}$ to the variables $y_{1,}\delta
y_{1,...,}\delta y_{N-1}$ one gets:%
\begin{eqnarray}
Z &=&\int\limits_{-\Delta }^{\Delta }D^{N-1}\delta y\left(
L-\sum_{i=1}^{N-1}\sigma _{i}\right) ^{N}  \label{Z 1} \\
&&\times \theta \left( L_{\max }^{\prime }-\sum_{i=1}^{N-1}\sigma
_{i}\right) \theta \left( \sum_{i=1}^{N-1}\sigma _{i}-(N-1)\sigma
_{m}\right) .  \notag
\end{eqnarray}%
The $\theta $ functions restrict the $y$ integration domain to those $\{y\}$
for which $\sigma ^{\prime }s$ are in the allowed range, $\sigma _{m}\leq
\sigma _{j}(\delta y_{j})\leq 1$, but their sum does not exceed $L_{\max
}^{\prime }.$ However, rather than integrating over the entire $\delta y$
domain defined by the $\theta $ function, it is convenient first to fix the
condensate's length at some $L^{\prime }$ and then integrate over\ its
possible values. This can be done by introducing following representation of
the step function $\theta $: 
\begin{eqnarray}
&&\theta \left( L_{\max }^{\prime }-\sum \sigma _{j}\right) \theta \left(
\sum \sigma _{i}-(N-1)\sigma _{m}\right)  \label{Teta} \\
&=&\int_{(N-1)\sigma _{m}}^{L_{\max }^{\prime }}dL^{\prime }\delta \left(
L^{\prime }-\sum \sigma _{i}\right) ,  \notag
\end{eqnarray}%
where one diameter $d$ is neglected in comparison to $L.$ From now on we set 
$d$ $=1.$ Next we change the integration over $\delta y$ to that over $%
\sigma ,$ $d\delta y_{i}=d\widetilde{\sigma }_{i}=\sigma _{i}d\sigma _{i}/%
\sqrt{1-\sigma _{i}^{2}}.$ Then, in the context of (\ref{Teta}), $Z$ becomes%
\begin{equation}
Z=\int_{N\sigma _{m}}^{L_{\max }^{\prime }}dL^{\prime }(L-L^{\prime
})^{N}\int_{\sigma _{m}}^{1}D^{N-1}\widetilde{\sigma }\delta \left(
L^{\prime }-\sum \sigma _{i}\right) .  \label{Z2}
\end{equation}%
We use the analytical representation of the delta function in $Z:$ 
\begin{equation}
\delta \left( L^{\prime }-\sum \sigma \right) =\frac{1}{2\pi }%
\int\nolimits_{-\infty }^{\infty }d\alpha e^{i\alpha \left( L^{\prime }-\sum
\sigma \right) }.  \label{del fun}
\end{equation}%
Next we perform the $\widetilde{\sigma }$ integration which factorizes into $%
N-1$ similar integrals to obtain%
\begin{equation}
Z=\int\nolimits_{-\infty }^{\infty }d\alpha \int_{N\sigma _{m}}^{L_{\max
}^{\prime }}dL^{\prime }e^{i\alpha L^{\prime }}(L-L^{\prime })^{N}\left(
\int_{\sigma _{m}}^{1}d\widetilde{\sigma }e^{-i\alpha \sigma }\right) ^{N-1}.
\label{Z4}
\end{equation}%
It is convenient to introduce the per disk lengths $l^{\prime }=L^{\prime
}/N $, $l_{\max }^{\prime }=L_{\max }^{\prime }/N,l=L/N$ and rewrite the PF
in the following form: 
\begin{equation}
Z=\int_{\sigma _{m}}^{l_{\max }^{\prime }}dl^{\prime }\int d\alpha e^{Ns},
\label{Z3}
\end{equation}%
where the factor $N^{N}$ is omitted, $N-1$ is replaced by $N$ and 
\begin{equation}
s=i\alpha l^{\prime }+\ln (l-l^{\prime })+\ln \left( \int_{\sigma _{m}}^{1}d%
\widetilde{\sigma }e^{-i\alpha \sigma }\right) .  \label{S}
\end{equation}%
\qquad Now we can compute the PF (\ref{Z3}) by the steepest descent method.
It is convenient to introduce $a=i\alpha $ which is real since $\alpha $ at
the saddle point lies on the imaginary axis and the integration contour has
to be (and can be) properly deformed,\ Fig. 2. The saddle point in the limit 
$N\rightarrow \infty $ determines the integral (\ref{Z3}) exactly. It is
specified by the stationary point of the function $s$ (\ref{S}) which, for
given $l$ and $\sigma _{m},$ depends on $a$ and $l^{\prime }.$ The two
equations $\partial s/\partial a=\partial s/\partial l^{\prime }=0$ can be
reduced to the single equation which reads: 
\begin{eqnarray}
\int\nolimits_{\sigma _{m}}^{1}d\sigma f_{\sigma }(\sigma ,a)\sigma &=&%
\overline{\sigma },  \label{l} \\
\overline{\sigma } &=&l-1/a,  \label{s}
\end{eqnarray}%
where we introduced the function%
\begin{equation}
f_{\sigma }(\sigma ,a)=\frac{\sigma e^{-a\sigma }}{\sqrt{1-\sigma ^{2}}%
\int_{\sigma _{m}}^{1}\frac{d\sigma \sigma e^{-a\sigma }}{\sqrt{1-\sigma ^{2}%
}}}.  \label{f}
\end{equation}%
\begin{figure}[tbp]
\includegraphics[height=7cm]{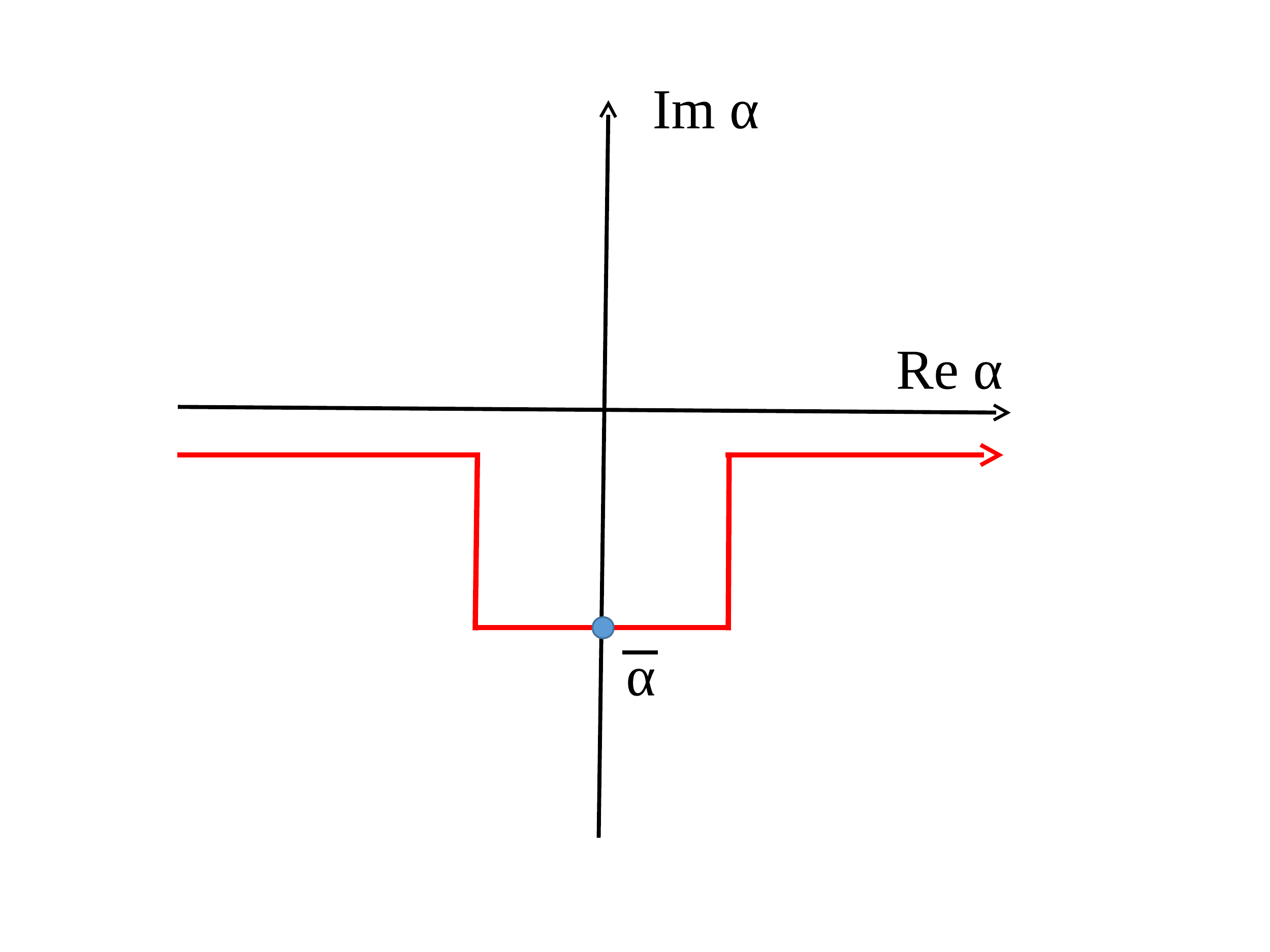}
\caption{{}The intergation contour over $\protect\alpha $ deformed as to
pass through the saddle point $\overline{\protect\alpha }.$ The length of
the segment with this point is on the order of $1/\protect\sqrt{N}.$}
\label{Fig2}
\end{figure}
The solution $\overline{a}$ of equation (\ref{l}) depends on the per disk
pore length $l$ and, via $\sigma _{m},$ on the pore width $D,$ and fully
determines the free energy. The free energy $F$ per disk, which therefore is
the function of the length $l$, width $D,$ and the temperature~$T$, is $%
F(l,D,T)=-Ts(\overline{a})=-TS$ where $S$ is system's per disk entropy (up
to terms independent of $L$ and $D$):%
\begin{equation}
S=\overline{a}\overline{\sigma }+\ln (l-\overline{\sigma })+\ln \left(
\int_{\sigma _{m}}^{1}\frac{d\sigma \sigma e^{-\overline{a}\sigma }}{\sqrt{%
1-\sigma ^{2}}}\right) .  \label{S1}
\end{equation}%
Finally, the PF has the form%
\begin{equation}
Z=\exp (NS),  \label{ZZZ}
\end{equation}%
where the prefacfor $\sim \sqrt{N}$ is omitted since in the thermodynamic
limit its contribution is negligible in comparison to $NS.$

The saddle point\ $(i\overline{\alpha }=\overline{a}>0,l^{\prime }=\overline{%
\sigma })$ determined by the equations (\ref{l}) and (\ref{s}) is the point
of maximum of the function $s(i\alpha ,l)$ on the integration contour shown
in Fig. 2 as the quadratic form $\delta ^{2}s$ at this point is negative:%
\begin{eqnarray}
\delta ^{2}s(i\alpha &=&\overline{a},l^{\prime }=\overline{\sigma })
\label{d2s} \\
&=&-(l-\overline{\sigma })^{2}(l^{\prime }-\overline{\sigma })^{2}-(%
\overline{\sigma ^{2}}-\overline{\sigma }^{2})(Re\alpha )^{2}\leq 0,  \notag
\end{eqnarray}%
where $\overline{\sigma ^{2}}=\int d\widetilde{\sigma }f_{\sigma }\sigma
^{2}.$ As to uniqueness of the saddle point, equation (\ref{s}) and the
relation between $1/a$ and the pressure, eq.(\ref{PL}) below, show that\ a
physically acceptable $\overline{\alpha }$ must lie on the negative
imaginary axis so that solutions to the equation (\ref{l}) should be sought
for real positive $a$. Our numerical findings have shown no sign of two
different real positive solutions to equation (\ref{l}) and in what follows
we will assume that the solutions presented for different $l$ and $D$ are
unique.

Note that the method applied above for solving the PF integral essentially
consists in making use of an explicit analytical representation of the
singular functions that determine the integration domain first in $\delta x$
and $y$, then in $\delta y,$ and finally in $\sigma .$ In Appendix A, I show
however that integration order over coordinates and the quantity fixing the
integration domain, e.g., $L^{\prime }$ or $L,$ is important and changing it
has to be made with circumspection.

\section{The pressures}

The q1D system is anisotropic and has two different pressures: the
longitudinal, 
\begin{equation}
P_{L}=T(\partial S/\partial l)_{D}/D,  \label{PPL}
\end{equation}%
and the transverse 
\begin{equation}
P_{D}=T(\partial S/\partial D)_{l}/l.  \label{PPD}
\end{equation}

The $l$ differentiation in (\ref{PPL}) can be readily performed. The
derivative $\partial $ $\overline{\sigma }/\partial l$ can be computed as
the $l$ derivative of the r.h.s. of equation (\ref{l}). After a simple
algebra one obtains:%
\begin{equation}
P_{L}=\frac{T}{D(l-\overline{\sigma })}.  \label{PL}
\end{equation}

The $D$ differentiation in (\ref{PPD}) can also be readily obtained.
Regarding for the relation (\ref{sigma}) between $\sigma _{m},$ $\Delta ,$
and $D,$ one obtains%
\begin{equation}
P_{D}=\frac{T\exp \left( -\frac{\sigma _{m}}{l-\overline{\sigma }}\right) }{%
ld\int_{\sigma _{m}}^{1}\frac{d\sigma \sigma e^{-\overline{a}\sigma }}{\sqrt{%
1-\sigma ^{2}}}.}  \label{PD}
\end{equation}%
If the length per disk is replaced by the linear density $\rho =N/L=1/l,$
the pressures can be expressed as functions of $\rho :$%
\begin{eqnarray}
P_{L} &=&\frac{\rho T}{D(1-\rho \overline{\sigma })},  \label{PP} \\
P_{D} &=&\frac{\rho T\exp \left( -\frac{\rho \sigma _{m}}{1-\rho \overline{%
\sigma }}\right) }{d\int_{\sigma _{m}}^{1}\frac{d\sigma \sigma e^{-\overline{%
a}\sigma }}{\sqrt{1-\sigma ^{2}}}.}  \notag
\end{eqnarray}

These are remarkably simple formulas. While anticipating the form of $P_{D}$
from the general ideas is hardly possible, the formula for $P_{L}$ is very
natural and could be expected from the starting expression for the PF (\ref%
{Z 1}). The PF of a one-dimensional HD system obtained by Tonks is $%
(L-Nd)^{N}$ and the pressure, in our notations, is $\sim (l-d)^{-1}.$ But
the formula (\ref{Z 1}) is \ the value of $(L-\sum \sigma )^{N}$ averaged
over the distribution of $\sigma $ and can be expected to give something
like $(L-N\overline{\sigma })^{N}\propto (l-\overline{\sigma })^{N}$ so that 
$P_{L}$ is naturally expected to scale as $(l-\overline{\sigma })^{-1}$
which is indeed the exact result (\ref{PL}). Moreover, in two limiting cases
of a very small density and density at the close packing, the pressure along
the pore has to coincide with the modified Tonks result $P_{L}\sim (l-\sigma
_{m})^{-1}$ in which $d$ is replaced with the minimum possible distance
along the pore. Indeed, in the former limit, it is because the pressure must
scale as $1/l$; in the last limit, it is because the motion across the pore
is fully hindered so that $\sigma _{m}$ plays the role of $d.$ The pressure
in the form (\ref{PL}) naturally satisfies these requirements.

The quantity $\overline{\sigma }$ is a smooth function of both density $\rho 
$ and thickness $D.$ In particular, $\overline{\sigma }\rightarrow d$ as $D$
goes to zero. Then the pressure $P_{L}$ is an analytical function of $\rho $
up to the close packing density $\rho =1/\sigma _{m}.$ This shows that the
virial expansion of $P_{L}$ does exist and converges to the exact pressure
for all $\rho <1/\sigma _{m}.$ In particular, the virial series times $D$
converges to the 1D pressure. Similarly, the transverse pressure $P_{D}$ can
also be expanded in a power series of $\rho $ which is convergent up to the
close packing density. The virial expansion of $P_{D}$ though has not been
considered so far.

The expressions for $P_{L}$ and $P_{D}$ which are similar in their form to
our results (\ref{PL}) and (\ref{PD}) were obtained by Wojciechovski and
coworkers in Ref.\cite{Wojc} in which the PF was also derived by finding a
stationary point of some functional. The direct comparison with our result
is however difficult as these authors considered a q1D system periodic in
the $y$ direction. Moreover, there is a more substantial difference between
this and our appraoch which is addressed in Appendix A.

\subsection{Distribution of disks' centers across the pore.}

We show here that the function $f_{\sigma }$ (\ref{f}) gives the probability
distribution of contact distances $\sigma $ along the pore hence equation (%
\ref{l}) gives its mean value $\overline{\sigma }.$ Clearly, in view of the
relation (\ref{sigma}), $f_{\sigma }$ also determines the distribution of $%
\delta y$ which is the difference between the $y$ coordinates of two
neighbors. We can also derive an analytical formula for the distribution
function of $y$, the coordinates of disks' centers across the pore $f_{y}.$
This probability distribution is the mean value of the $y$ coordinate of a
single disk, say disk $k,$ i.e., $\left\langle \delta (y-y_{k})\right\rangle
.$ To find it, we fix the transverse coordinate of $k$-th disk, $y_{k}=y>0,$
and separate the $\delta y$ integrals over $\delta y_{k-1}$ and $\delta
y_{k} $ from the rest \thinspace $N-3$ $y$ integrals in formula (\ref{Z 1}):%
\begin{eqnarray}
\left\langle \delta (y-y_{k})\right\rangle &\propto &\int\limits_{-\Delta
}^{\Delta }D^{N-1}\delta y\delta (y-y_{k})  \label{dy} \\
&=&\left[ \int\limits_{-\Delta /2-y}^{\Delta
/2-y}d(y-y_{k-1})\int\limits_{-\Delta /2-y}^{\Delta /2-y}d(y_{k+1}-y)\right]
\notag \\
&&\times \int\limits_{-\Delta }^{\Delta }D^{N-3}\delta y.  \notag
\end{eqnarray}

Changing to the integration over $\sigma ,$ this reads%
\begin{equation}
\int_{\sigma _{m}}^{1}D^{N-1}\widetilde{\sigma }\delta
(y-y_{k})=\int_{\sigma _{m}}^{1}D^{N-3}\widetilde{\sigma }\times
I_{k}I_{k-1},  \label{d sigma}
\end{equation}%
where%
\begin{equation}
I_{k}=\left[ \int\limits_{\sqrt{1-(y-\Delta /2)^{2}}}^{1}d\widetilde{\sigma }%
_{k-1}+\int\limits_{\sqrt{1-(y+\Delta /2)^{2}}}^{1}d\widetilde{\sigma }_{k-1}%
\right] .  \label{Ikk}
\end{equation}%
Next we separate terms in the delta function in (\ref{del fun}) that depend
on $\sigma _{k-1}$ and $\sigma _{k}$; the sum $\sum \sigma $ is not affected
as removing two links of finite length out of $N-1$ links is negligible in
the thermodynamic limit. The probability of $k$-th disk to stay at $y$ is
proportional to the PF in the form given in (\ref{Z2}) where $\int_{\sigma
_{m}}^{1}D^{N-1}\widetilde{\sigma }$ is replaced by (\ref{d sigma}). As in
such expression the only $y$ dependent terms are $I_{k}I_{k-1}\exp (-a\sigma
_{k}-a\sigma _{k-1}),$ we conclude that the distribution function of $y$
normalized on unity is of the form%
\begin{eqnarray}
f_{y}(y) &=&\left[ \int\limits_{\sqrt{1-(y-\Delta /2)^{2}}}^{1}\frac{\sigma
e^{-\overline{a}\sigma }d\sigma }{\sqrt{1-\sigma ^{2}}}+\int\limits_{\sqrt{%
1-(y+\Delta /2)^{2}}}^{1}\frac{\sigma e^{-\overline{a}\sigma }d\sigma }{%
\sqrt{1-\sigma ^{2}}}\right] ^{2}  \notag \\
&&\times \frac{1}{A},  \label{fyy} \\
A &=&\int_{-\Delta /2}^{\Delta /2}dyf_{y}A.  \notag
\end{eqnarray}%
Here $\overline{a}$ depends on $l$ and $D$ as the solution of equation (\ref%
{l}). As this expression is symmetric with respect to the sign of $y$, there
is no need to consider the case $y<0$ separately.

In a similar way, one can show that the function $f_{\sigma }$ (\ref{f})
with $a=\overline{a}$ is the distribution function of $\sigma .$ This
probability distribution is the mean value of $\sigma $ for some $k$, i.e., $%
\left\langle \delta (\sigma -\sigma _{k})\right\rangle .$ We fix $\sigma
_{k} $ at $\sigma $ in\ the integrand of the integral over $\sigma _{k}$ in (%
\ref{Z 1}) which gives the expression in the nominator of formula (\ref{f}).
Normalizing on unity, one obtains $\left\langle \delta (\sigma -\sigma
_{k})\right\rangle =f_{\sigma }.$ Both analytically obtained distribution
functions $f_{\sigma }(\sigma )$ and $f_{y}(y)$ will be presented below for
different parameters of the pore.

\section{Pressure and disks' arrangement in the pore}

The pressures\ along and across the pore as functions of the linear density $%
\rho =N/L$ are presented for three different pore widths: $\Delta =0.141$
close to the 1D case$,$ $0.5,$ and $\sqrt{3}/2\approx 0.866,$ the maximum
width in the q1D system; the quantities $P_{L}$ and $P_{D}$ are shown in
Fig.3 where also presented is the contribution of the term $1/D(l-\sigma
_{m})$. Consider the transverse pressure $P_{D}$. We see that for low
densities this pressure is higher than the one along the pore. This is
because in that case $P_{L}$ is determined by a large $x$ disks' separation
whereas $P_{D}$ is determined by a short range of $y$ motion which is
bounded from above by $\Delta .$ At sufficiently high density however the $x$
separation becomes comparable with $\Delta $ and the two pressures may
intersect. The disks' separation $l-\overline{\sigma }$ along $x$ is $%
1/(P_{L}D).$ At the crossing points in Figs.3a and 3b one finds that for $%
\Delta =0.141,$ $l-\overline{\sigma }=0.13,$ and for $\Delta =0.5,$ $l-%
\overline{\sigma }=0.33$ which are slightly lower than the correspondent $%
\Delta $'s. This difference is because the disks mountain one upon another
thereby decreasing the range of $y$ motion and the stronger so the wider the
pore is. This qualitatively explains why the difference between $\Delta $
and $l-\overline{\sigma }$ at the pressure crossing point increases with the
pore width. For higher $\Delta ,$ this effect is so strong that disks' $y$
motion is highly restricted by their next neighbors and, e.g., for $\Delta
=0.86$ the transverse pressure is always above the longitudinal one. Both $%
P_{L\text{ }}$ and $P_{D}$ tend to the curve $1/(l-\sigma _{m})D$ in the
close packing density limit as it should be. Some more delicate
peculiarities of the pressure behavior are related with the appearance of
certain defects in the zigzag structure and will be discussed below in the
Discussion. 
\begin{figure}[tbp]
\includegraphics[width=10cm]{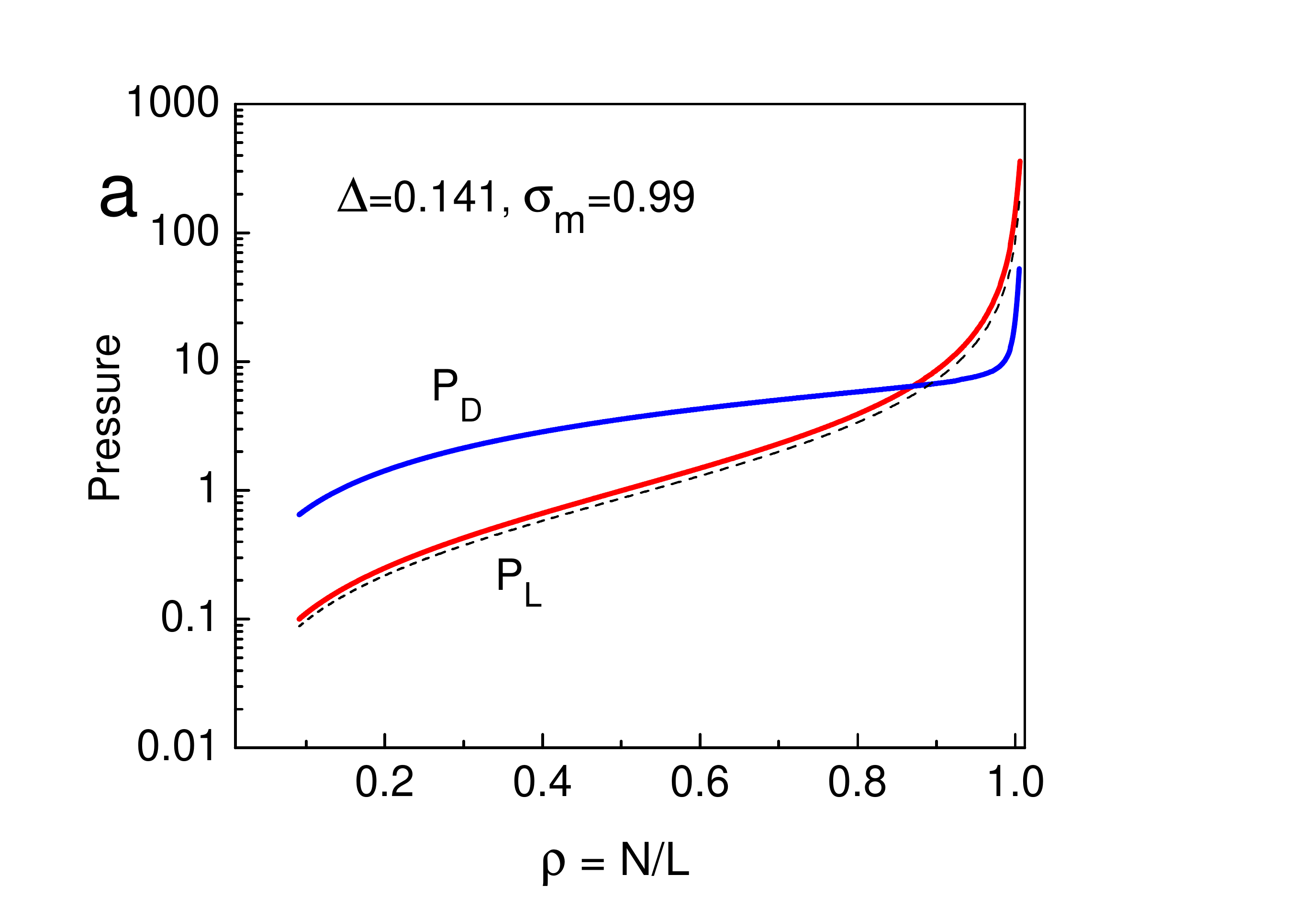} %
\includegraphics[width=10cm]{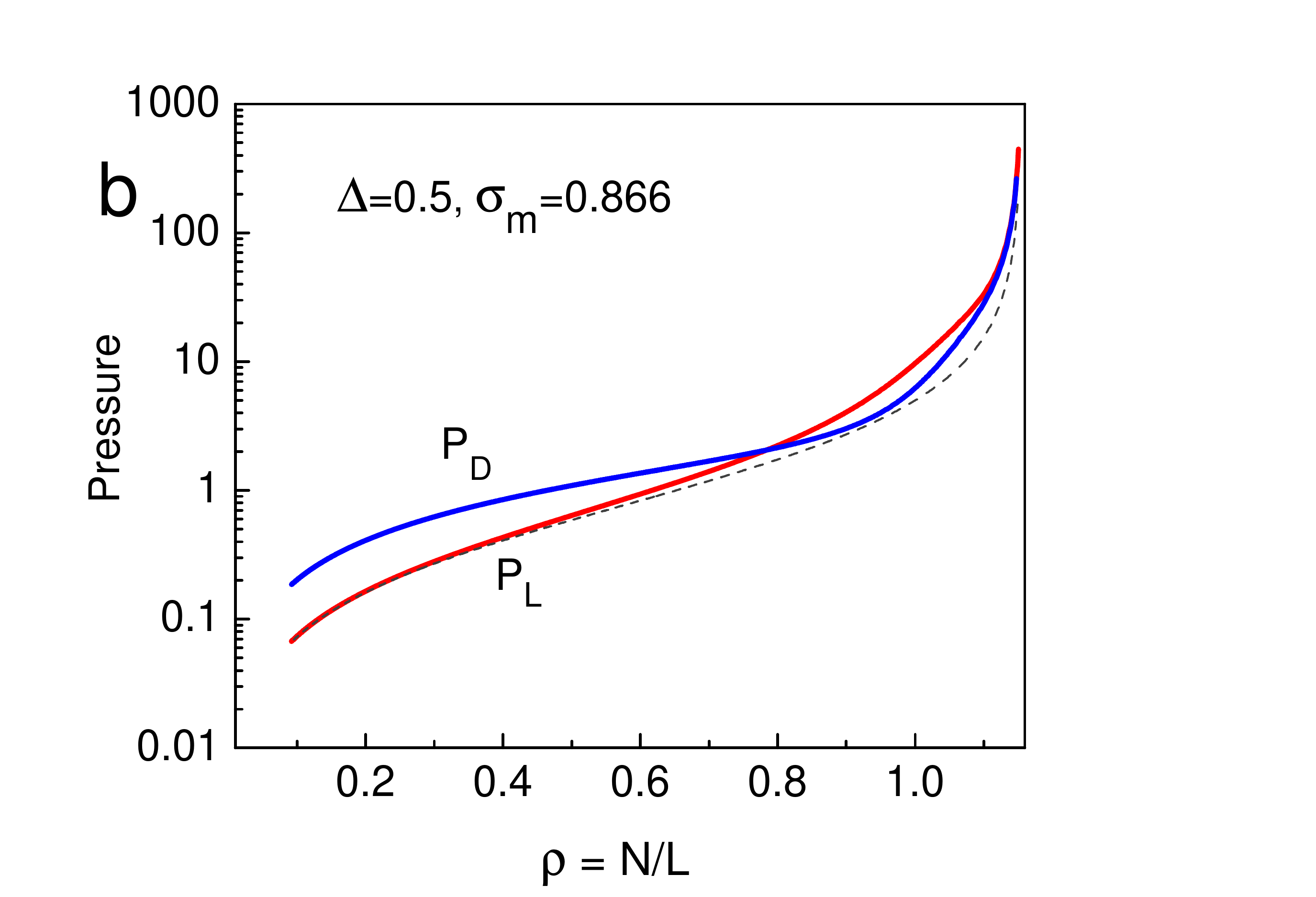} %
\includegraphics[width=10cm]{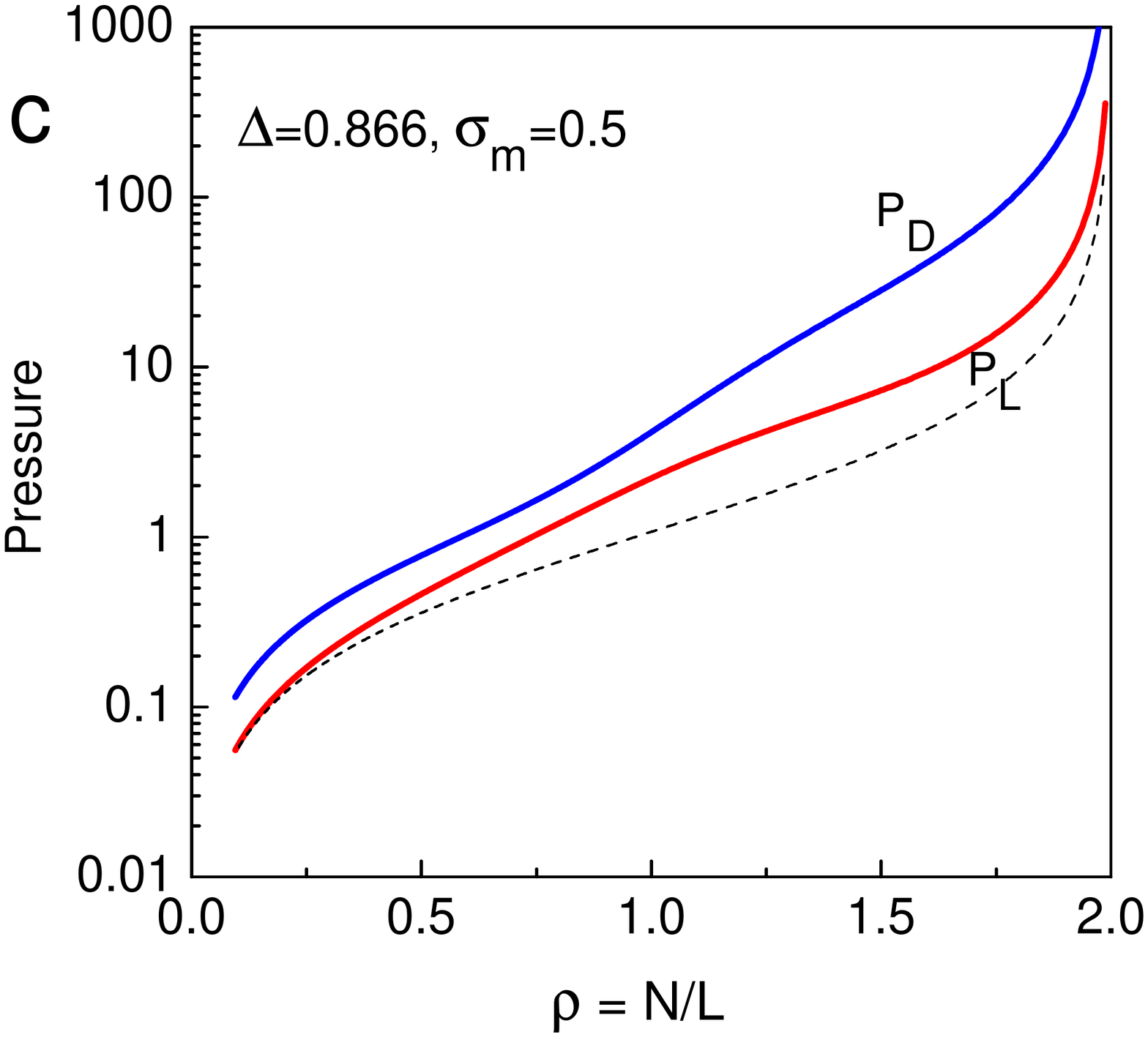}
\caption{The longitudinal, $P_{L},$ and transverse, $P_{D},$ pressures for
three different widths $\Delta $: a) 0.141, b) 0.5, c) 0.866. The dash
curves show the contribution of the term $1/[D(l-\protect\sigma _{m})]$ with
the relevant $\protect\sigma _{m}.$ $T$ is set equal to $1.$ Note that, in
Fig.3b, for large $\protect\rho $ the curve $P_{L}$ lies above $P_{D}$ so
that there is no second crossing.}
\label{Fig3}
\end{figure}

The function $f_{\sigma }(\sigma )$ (\ref{f}) with $a=\overline{a}$ presents
the distribution of the longitudinal contact distances $\sigma ,$ eq.(\ref%
{sigma}), and eq.(\ref{l}) gives its mean value $\overline{\sigma }.$ This $%
\overline{\sigma }$ is growing with $l$ and for $l\sim 3$ practically
attains its maximum limiting value $\overline{\sigma }_{\infty }=\overline{%
\sigma }(l\rightarrow \infty ).$ The limiting value $\overline{\sigma }%
_{\infty }$ is larger for smaller $\Delta $ but remains below $1$ for all $%
\Delta >0$ ($\overline{\sigma }_{\infty }=0.853,0.956,0.995$ for $\Delta
=0.866,0.5,0.141$ respectively$),$ and only for $\Delta =0,$ i.e., in the 1D
case, $\overline{\sigma }=\overline{\sigma }_{\infty }=1.$ This shows that
the piecewise discontinuity in $l^{\prime }$ at the upper $l^{\prime }$
integration limit $l_{\max }^{\prime }$ in (\ref{Z3}) is never attained and
thus does not manifest itself.

\begin{figure}[tbp]
\includegraphics[width=10cm]{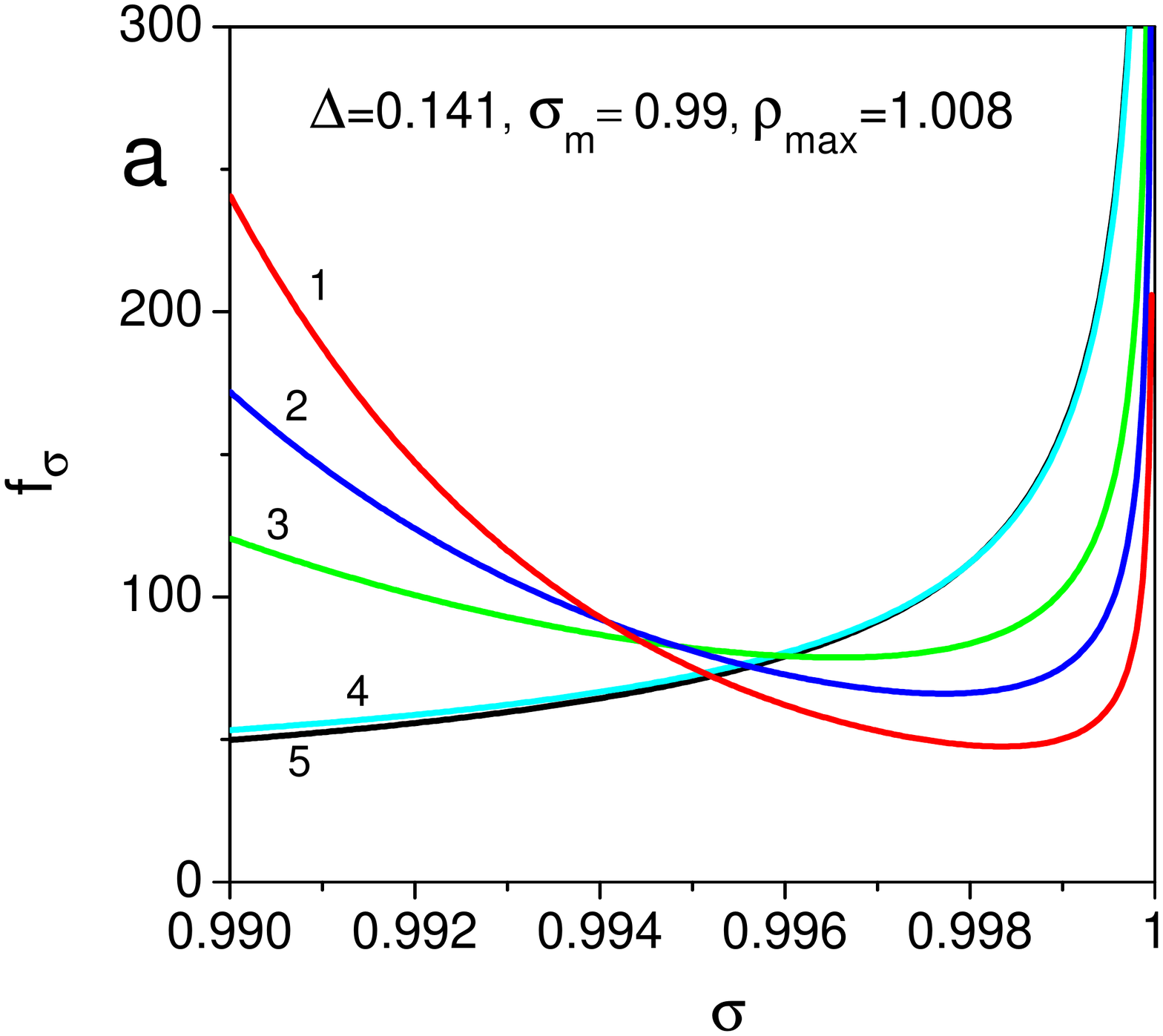} %
\includegraphics[width=10cm]{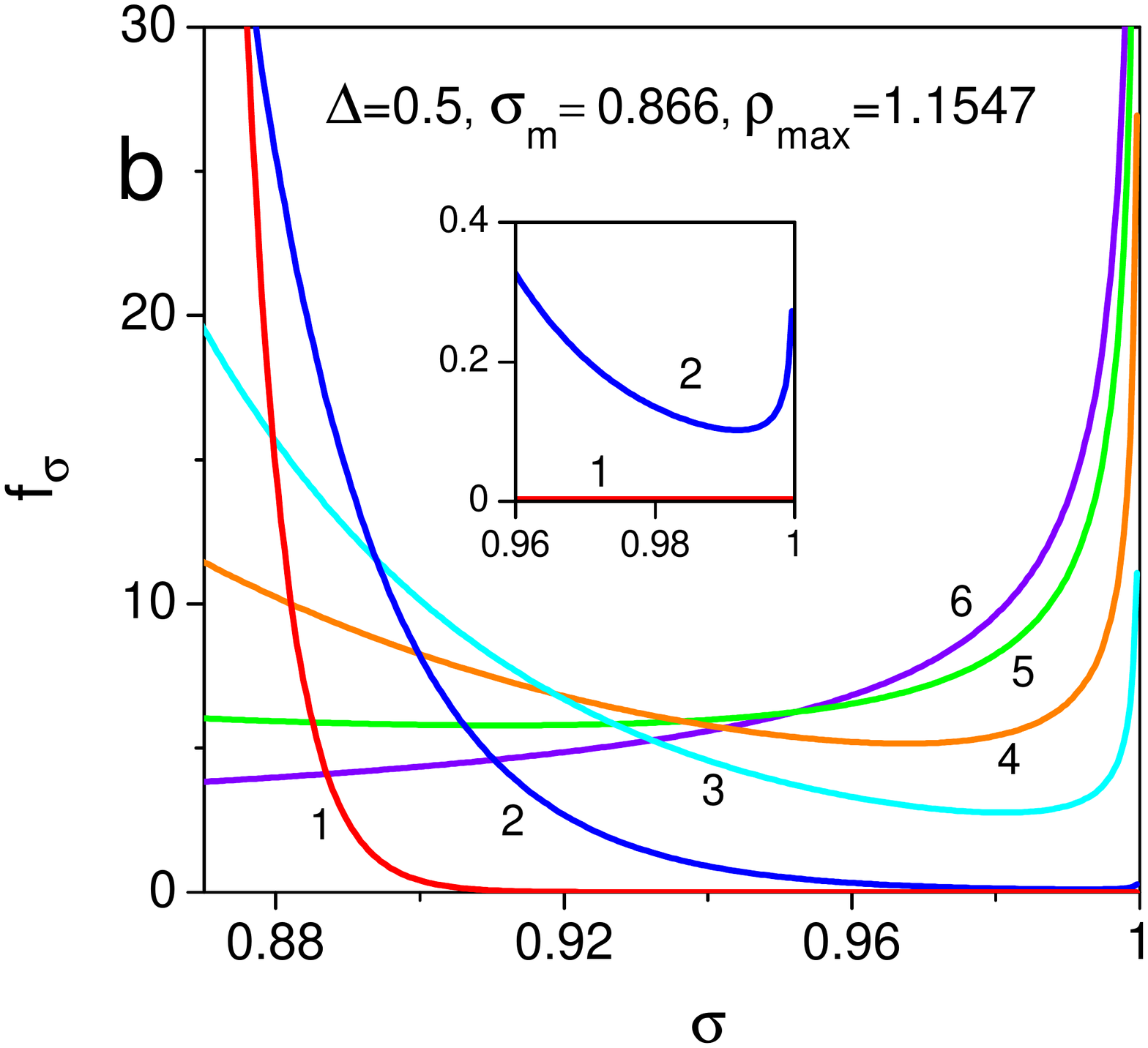} %
\includegraphics[width=10cm]{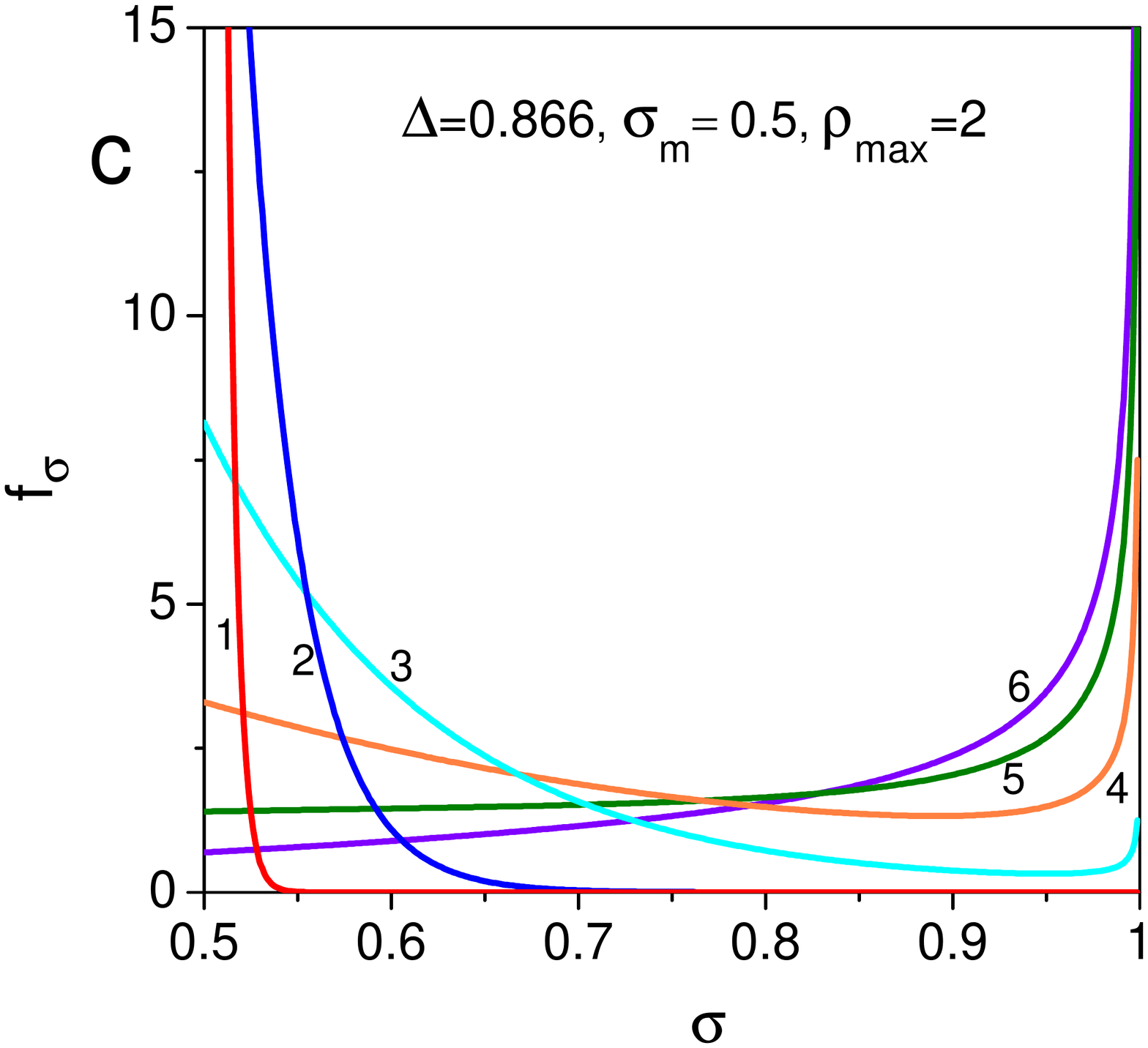}
\caption{The distribution $f_{\protect\sigma }$ of the contact distances $%
\protect\sigma $ in the condensates correspondent to three different widths $%
\Delta $ and various densities $\protect\rho :$ a) $\Delta =$0.0141: curve 1-%
$\protect\rho =1.005,$ 2-1.003, 3-1, 4-0.9, 5-0.091. b) $\Delta =$0.5:
1-1.14, 2-1.111, 3- 1.056, 4-1.01, 5-0.79, 6-0.5. Inset: curves 1 and 2 near 
$\protect\sigma =1.$ c) $\Delta =$0.866: 1-1.96, 2-1.8, 3-1.4, 4-1.1, 5-0.8,
6-0.1. }
\label{Fig4}
\end{figure}
\begin{figure}[tbp]
\includegraphics[width=9cm]{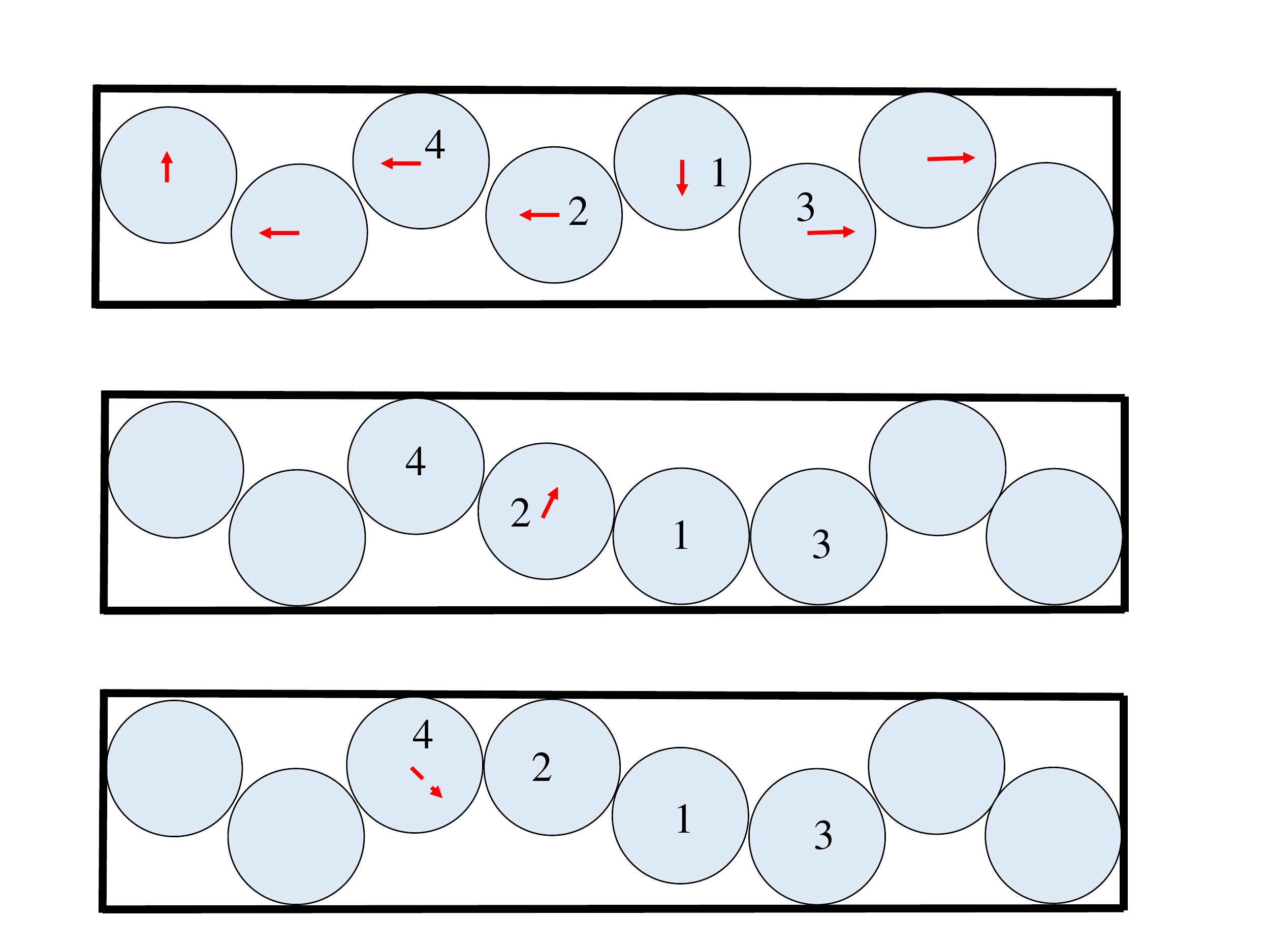}
\caption{{}Disks' rearrangement in a pore which creates a window for two
disks to exchange their vertical positions. Upper panel: disk in the pore at
the average distance along the pore which is below the diameter $d$ and
disks cannot exchange their vertical positions. To let disk 1 go down, disks
on the left and right of it get more dense. Mid panel: disk 1 gets down
through the window of size $d$ between disks 2 and 3. Now disk 2 may get up
between disks 4 and 1. Lower panel: the exchange of the vertical positions
of disks 1 and 2 is accomplished. Now disk 4 potentially can move down.}
\end{figure}

\begin{figure}[h]
\includegraphics[width=8cm]{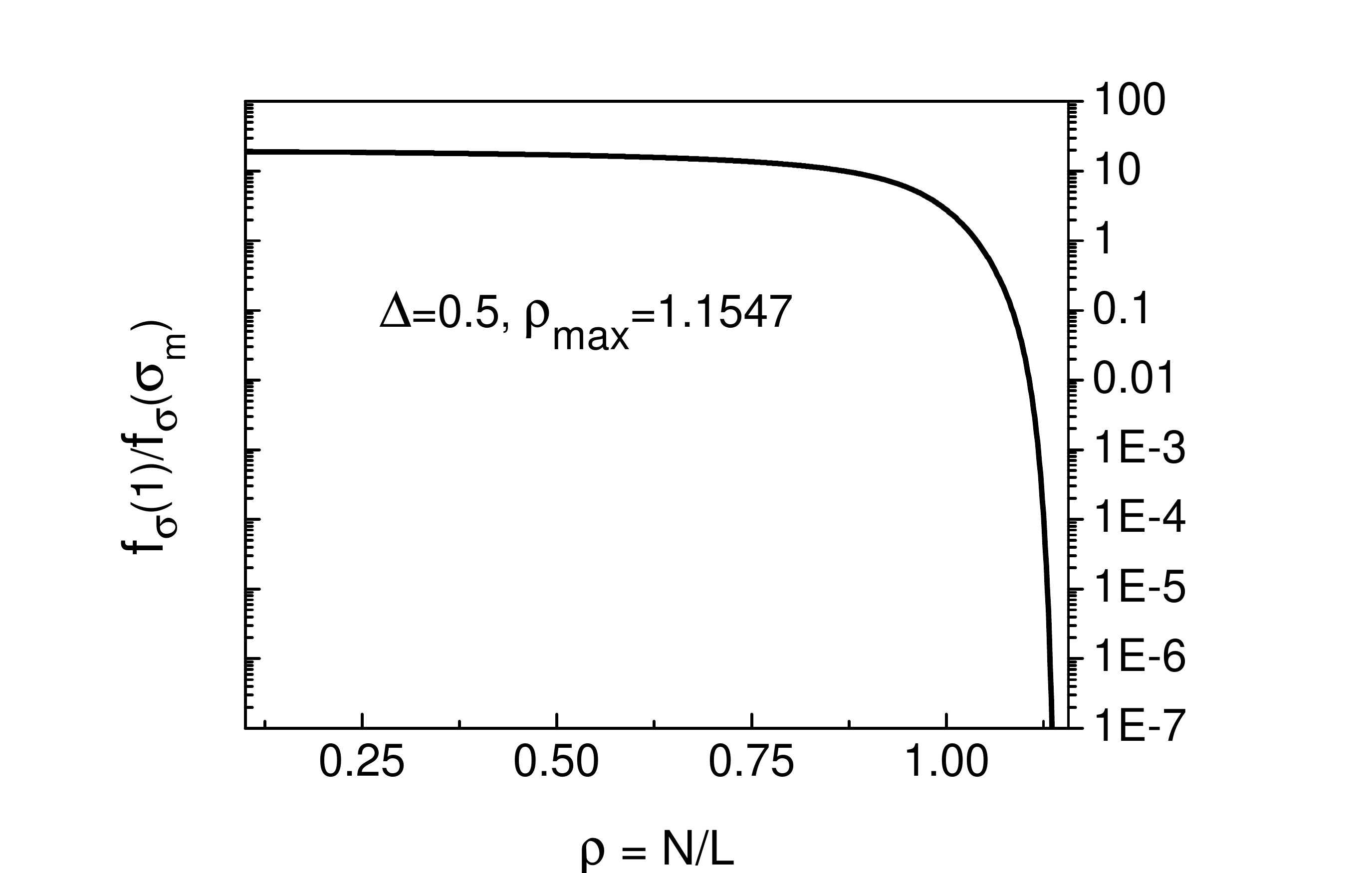}
\caption{{}Ratio $f_{\protect\sigma }(\protect\sigma =1)/f_{\protect\sigma }(%
\protect\sigma =\protect\sigma _{m})$ as a function of the linear density $%
\protect\rho $ for $\Delta =0.5.$ }
\label{Fig6}
\end{figure}

The distributions $f_{\sigma }(\sigma )$ for $\Delta =0.141,$ $0.5,$ $0.866$
is shown in Fig.4 for different densities $\rho =N/L$. Consider first $%
f_{\sigma }$ for the case $\Delta =0.5,$ Fig.4.b, recently studied
numerically by Huerta et al \cite{A i T,WE}. This $\sigma $ distribution has
an important peculiarity: it has two peaks, one at the smallest $\sigma
=\sigma _{m}$ and another one at the largest $\sigma =1,$ and a flat minimum
in between. At a large density $\sigma _{m}$ dominates implying that disks
contact the opposite walls making a solidlike zigzag. At the same time, $%
\sigma =1$ indicates that some disks can move across the pore through
windows between the zigzags, Fig. 5. The second peak appears quite sharply
in terms of the density variation, but not abruptly: it is present for any $%
\rho $, Fig.6, but becomes barely visible at about $\rho \approx 1.111$
(inset in Fig.5b) and well developed at $\rho =1.056.$ For $\rho $ $>1.111$
the $\sigma _{m}$ peak dominates, at $\rho \sim 1.056$ the $\sigma =1$ peak
becomes well visible$,$ then it grows and for $\rho <1$ becomes higher than $%
\sigma _{m}$ peak. This implies that at $\rho $ $\sim 1.06$ an appreciable
fraction of the zigzag arrangement is replaced by strings of disks with
close $y$'s. At lower $\rho <1$, the $\sigma $ distribution becomes wide
which shows that at low $\rho $ disks freely move between the walls as in an
ideal gas. This picture is in a qualitative agreement with the numerical
simulation results of Refs.\cite{A i T,Varga,WE}. The pair distribution
function along the pore was found to have sharp peak at the contact distance 
$\sigma _{m}$ at high density $\rho =1.11$, then it widens and, for $\rho
\approx 1.056,$ develops second peak at the unit distance which then widens
and becomes dominating for $\rho =0.91.$ The system behavior for $\Delta
=0.866$ is qualitatively similar to that of $\Delta =0.5.$ As for $\Delta
=0.141$, this width is so narrow and the positions at the wall and at the
center are so close that both peaks are present for the density 1.005
extremely close to the dense packing limit 1.008.

Fig. 7 presents the distribution function $f_{y}$ (\ref{fyy}) of coordinates 
$y$ across the pore. It is shown for three pore widths and different
densities up to very high ones close to the maximum possible dense packing
densities.

\begin{figure}[tbp]
\includegraphics[width=10cm]{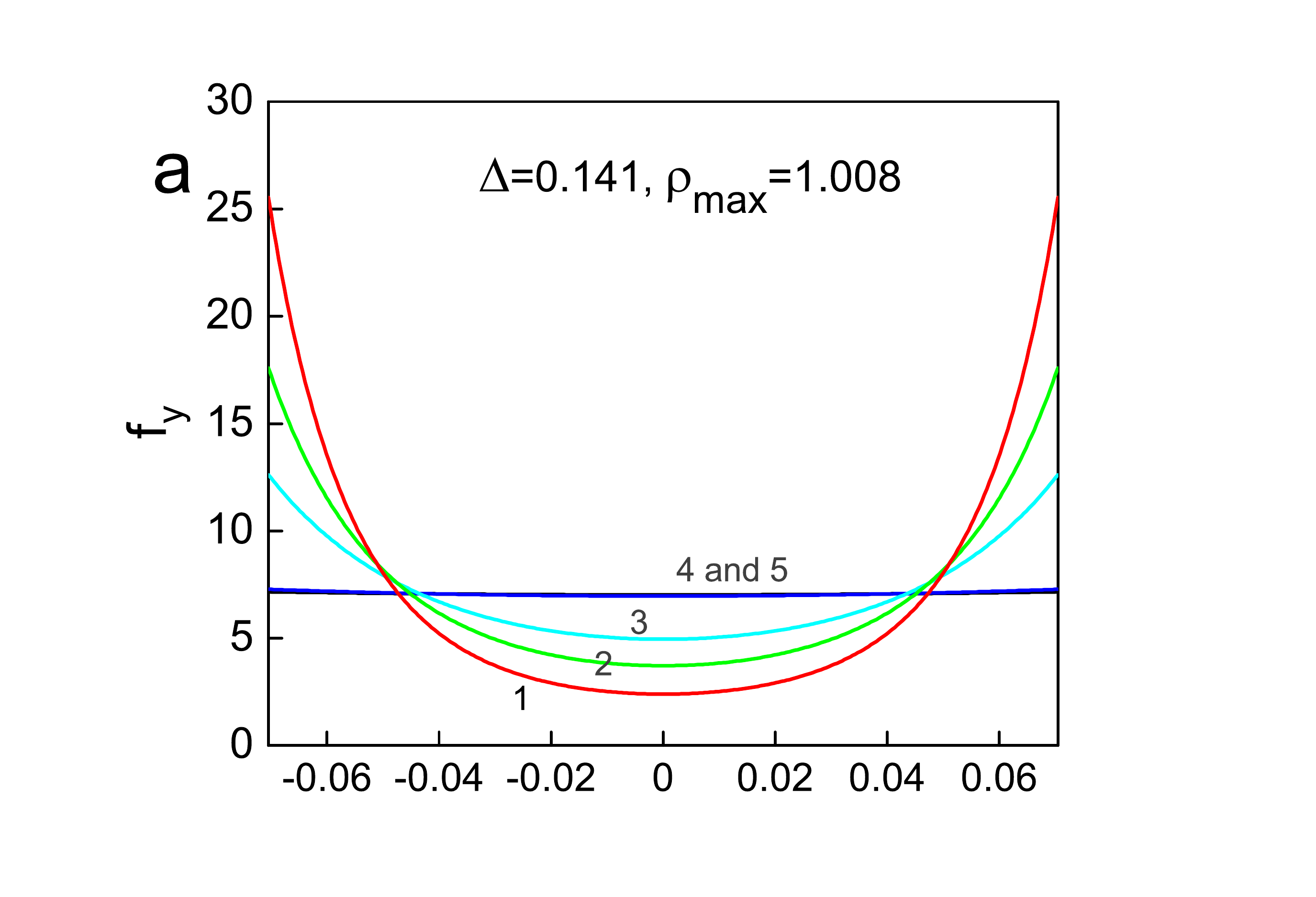} %
\includegraphics[width=10cm]{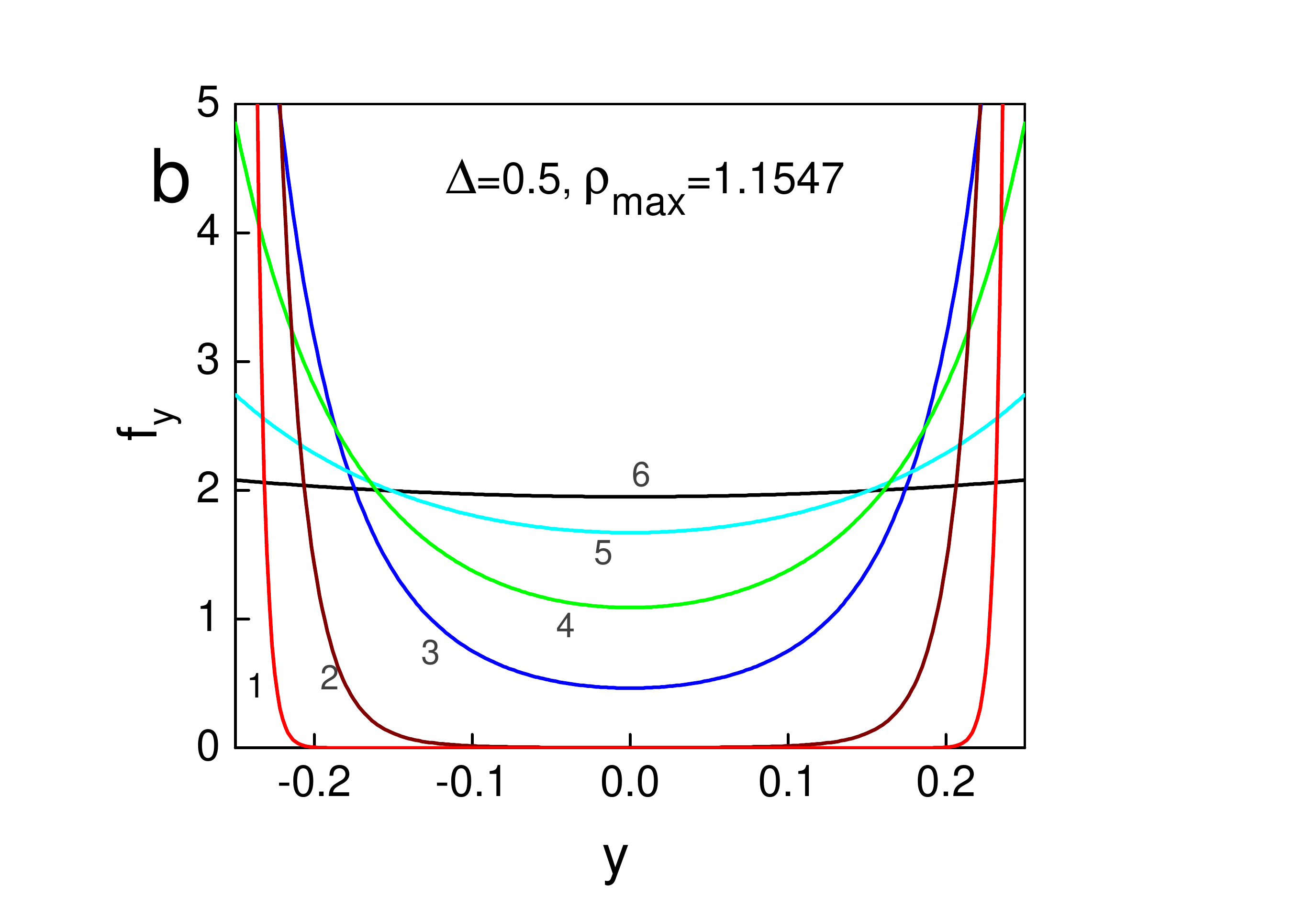} %
\includegraphics[width=10cm]{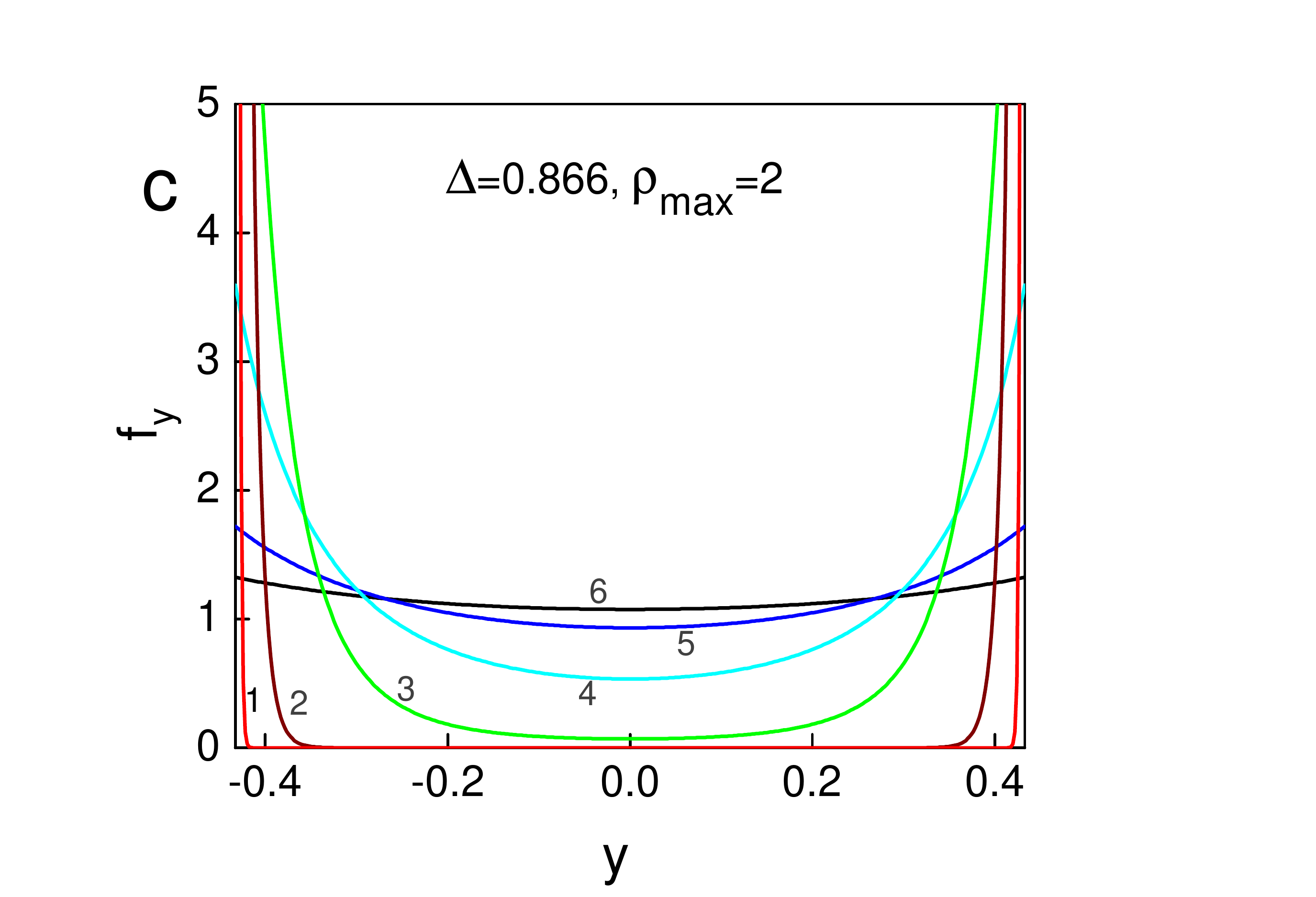}
\caption{The distribution function $f_{y}$ of the disk coordinate $y$\
across the pore for three pore widths $\Delta $ and different densities $%
\protect\rho $ : a) $\Delta =0.0141:$ curve $1$-$\protect\rho =$1.005, 2-$%
1.003,$ 3-1, 4-0.9, 5-0.8. b) $\Delta =0.5,$ 1-1.14, 2-1.111, 3-1.056,
4-1.01, 5-0.79, 6-0.5. c) $\Delta =0.866,$ 1-1.96, 2-1.8, 3-1.4, 4-1.1,
5-0.8, 6-0.5. $\protect\rho _{\max }$ is the maximum dense packing density. }
\label{Fig6}
\end{figure}

\subsection{The 1D limit}

It is important to see how the results obtained for a q1D system behave
approaching a 1D system, i.e., in the limit $D\rightarrow 0,$ when $\sigma
_{m}\rightarrow 1$ ($\sigma _{m}\rightarrow d)$ as $\Delta \rightarrow 0.$
To this end, we first estimate the $\sigma $ integral in this limit:%
\begin{equation}
\int_{\sigma _{m}(\Delta )}^{1}\frac{d\sigma \sigma e^{-\overline{a}\sigma }%
}{\sqrt{1-\sigma ^{2}}}=e^{-\overline{a}}\Delta +O(\Delta ^{2}).  \label{I0}
\end{equation}%
Then the PF (\ref{ZZZ}) goes over into the following expression:%
\begin{equation}
Z(D\rightarrow 0)\approx (L-Nd)^{N}\Delta ^{N},  \label{ZZ1D}
\end{equation}%
which shows that in this limit the longitudinal and transverse degrees of
freedom factorize. The longitudinal pressure times $D$ in this limit
recovers its 1D form $\propto T/(l-1)$ while the transverse pressure takes
the form of that of an ideal gas in the volume $\Delta d,$ $P_{D}=T\rho
/\Delta d.$ The inhomogeneous ditsribution $f_{y}$ of the coordinates $y$
across the pore in this limit behaves like 
\begin{equation}
f_{y}\approx \frac{1+\overline{a}\Delta ^{2}\widetilde{y}^{2}/2}{\Delta },
\label{fy0}
\end{equation}%
where $0\leq \widetilde{y}\leq 1.$ Thus, the inhomogeneity amplitude
vanishes as pore width in power two. This is in line with the results of
Refs.\cite{Fran1,Fran2} where similar dependence on the slit thickness was
obtained by means of a perturbative approach to the transition from quasi 2D
HD system in a slit to pure 2D HD system.

\section{Discussion}

As we said above, a very small and extremely narrow peak at $\sigma =1$
exists at any, even very large density, Figs.4,6. The above picture suggests
that this peak is an essential part of the equilibrium state. At large $\rho 
$, but sufficiently decreased from the dense packing value, the disks choose
to move closer to the walls to get compressed into solidlike zigzag array
with\ the interparticle distance somewhat smaller than its average and $%
\sigma $ smaller than $\overline{\sigma }$ in order to provide windows with $%
\sigma $ close to unity (i.e., of size of the disk diameter), Fig.5. Through
these windows the disks can interchange their vertical positions, extend
their wondering to the total pore width and bring some entropy gain to the
whole system. The two HDs in the window form a bound pair: the disks roll
over each other's surface and their positions are highly correlated \cite%
{Exchange}. A local density increase needed to provide a window of size $%
\sigma =1$ implies some increase of the pressure along the pore. At the same
time, at such window the $y$ range of disk motion should somewhat widen
implying some decrease of transverse pressure. We interpret the slight
upturn in $P_{L}$ and slight downturn in $P_{D}$ at $\rho $ above $\rho =1$
in Fig.3c as a manifestation of this effect: at the density \ 1.1, Fig.4c,
the peak at $\sigma =1$ becomes well developed implying that the number of $%
\sigma =1$ windows is appreciable and can affect the pressures. As the
density drops, the correlation between the disks weakens, the pair
dissociates into free disks which can travel across the pore independently,
their number rises while the number of HDs at the walls diminishes, Fig.6.
This picture invokes similarity with a continuous Kosterlitz-Thouless
transition from solidlike to liquidlike phase of a crystal. The similarity
is supported by the numerical findings by Huerta et al \cite{A i T,WE} that
in the case $\Delta =0.5$ above $\rho \sim 1.111$ the longitudinal pair
correlation drops as a power law whereas below this $\rho $ it drops
exponentially. Thus, our theory shows that the crossover between the
solidlike zigzag and the liquidlike intermittence of zigzag and string
arrangements is sharp in the scale of density variation, but continuous so
that the thermodynamic potentials of a q1D system of HDs do not have
discontinuities. The last conclusion is similar to that achieved by Varga et
al \cite{Varga} based on the numerical study of a q1D HD system. We
emphasize that the narrow peak at $\sigma =1$ for any density is the effect
which can be lost in a finite system: only an infinite system can provide a
window with $\sigma =1$ for whatever density as its size is negligible in
the limit $N\rightarrow \infty .$

\section{Conclusion}

Recently HDs in q1D geometry have received a great deal of interest and
there is an indication that it will last. The transfer matrix method by
Kofke and Post is on the way of incorporating\ wider pores where the
interaction includes more than one next neighbor \cite{Godfrey,Gurin
2015,Gurin 2017,Hu}. Moreover, HDs in q1D geometry are nowadays considered
in a wider aspect related to glass transitions and HDs' dynamics \cite%
{Robinson,Yamchi,Fu,Hicks,A i T, Godfrey2020}. Our result gives the direct
method to get the thermodynamics of a q1D HD system for given $\rho ,L,D$
which is required both for equilibrium and glassy states. The $\sigma $
distribution (\ref{f}) derived here suggests a novel quantitative analysis
of the solidlike-to-liquidlike transformation and has already resulted in
some new ideas \cite{WE}. The analytical formulas (\ref{PL}),(\ref{PD}) and\
(\ref{fyy}) allow one to find the pressures along and across the system,
disks' distribution across the pore, and pair correlations in a q1D HD
system (the work is in progress) without the need to solve additional
numerical problesms. The result complements recent studies of low and high
nonphysical dimensions which will hopefully advance our understanding of HD
systems in the dimensions 2 and 3.

\section{Data Availability Statement}

Data sharing is not applicable to this article as no new data were created
or analyzed in this study.

\textit{Acknowledgment.}%
\begin{acknowledgements}
I am highly endebted to A. Trokhymchuk for numerous enlightening discussions. The work was supported by
VC 202 from NAS of Ukraine and and NRFU Project 2020.01/0144.
\end{acknowledgements}

\appendix{}

\section{Appendix A. Making use of analytical representations of singular
functions and integration order in the statistical integrals}

In HD systems, disks' coordinates are not independent and the integration
limits are given by complex nontrivial expressions. Making use of a step
function $\theta $ in the statistical integrals with complicated integration
limits is very convenient as it can formally simplify these limits so that
the problem of solving the PF shifts to the integrating with singular
functions. In this paper we demonstrated this in the case of a q1D HD
system. In our integrals (\ref{Dx2}), (\ref{Z 1}), or (\ref{Z2}) the
arguments of the singular functions depend on the system length $L$ and
condensate length $L^{\prime }$. Solving the PF, we enjoyed constant
integration limits in the coordinates first integrating over the coordinates
and then over $L^{\prime }.$ One may naturally ask if the order can be
reversed. Moreover, there is often a need to integrate the expressions of
the form (\ref{Dx2}), (\ref{Z 1}), or (\ref{Z2}) with respect to the system
length $L$, in particular, for Laplace's transform in $L.$ The next
extremely simple example shows that first integrating over $L^{\prime }$ or $%
L$ and then over the coordinates can result in an inconsistency.

Consider the integral 
\begin{equation}
I=\frac{1}{2}\int_{0}^{L}dL\int_{0}^{1}dx\int_{0}^{1}dy\theta (L-x-y). 
\tag{A1}
\end{equation}%
This is an alternative expression for the integral 
\begin{equation}
I=\int_{0}^{L}dL\int_{0}^{L}dx\int_{0}^{L-x}dy=L^{3}/6.  \tag{A2}
\end{equation}%
The reason for the form (A1)\ with the theta function is that it is
convenient to deal with the \textit{simple coordinate independent
integration limits }in $x$ and $y$. Consider $I$ and, for simplicity, assume
that $L\leq 1$ (the analytical results differ for $L<1$ and $L>1).$ First
integrating (A2) over $xy$ and then over $L$ gives the correct result:%
\begin{equation}
I_{Lxy}=\int_{0}^{L}dL\int_{0}^{L}dx(L-x)=L^{3}/6.  \tag{A3}
\end{equation}%
This result can be reproduced by using $\theta $ function in the analytical
form (\ref{Tet}) and integrating (A1) in the same order: 
\begin{align}
I_{Lxy}& =\frac{1}{2}\int_{0}^{L}dL\int \frac{d\alpha }{2\pi i\alpha }%
e^{i\alpha L}\left( \frac{e^{-i\alpha L}-1}{i\alpha }\right) ^{2}  \tag{A4}
\\
& =\frac{1}{2}\int_{0}^{L}dLL^{2}=L^{3}/6.  \notag
\end{align}%
Reversing the order in (A1) with theta function in the form (\ref{Tet}) one
gets:%
\begin{equation}
I_{xyL}=\frac{1}{2}\int_{0}^{1}dx\int_{0}^{1}dy\int \frac{d\alpha }{2\pi
i\alpha }\left[ \frac{e^{i\alpha (L-x-y)}}{i\alpha }-\frac{e^{-i\alpha (x+y)}%
}{i\alpha }\right] .  \tag{A5}
\end{equation}%
The second term in the bracket gives zero as its exponent is negative. The
first term has a pole of the second order in $\alpha $ which gives%
\begin{align}
I_{xyL}& = & & \frac{1}{2}\int_{0}^{1}dx\int_{0}^{1}dy(L-x-y)\theta (L-x-y) 
\notag \\
& = & & \frac{1}{2}\left(
\int_{0}^{L}dx\int_{0}^{L-x}dy+\int_{0}^{L}dy\int_{0}^{L-y}dx\right)  \notag
\\
& \times & & (L-x-y)  \notag \\
& = & & L^{3}/6  \tag{A6}
\end{align}%
We see that the reverse order also gives the correct result, but if the
integral was manyfold, one would encounter a serios problem in integrating
over the coordinates and here is why. If the $L$ integration is performed 
\textit{after }that over $x$ and $y,$ the integration limits in $x$ and $y$
are\textit{\ independent} and the $x,y$ integration is trivial, see (A4). In
contrast, if the $L$ integration is performed\textit{\ before} that over $x$
and $y,$ the integration limits in $x$ and $y$ are\textit{\ not independent }%
because of the presence of $\ \theta (L-x-y),$ see (A6). Thus, on the $L$
integration one again arrives at the integral with coordinate dependent
integration limits so that the goal has not been achieved: the integration
over different coordinates cannot be performed independently. We emphasize
that omitting $\theta (L-x-y)$ in (A7) does give an integral with constant
integration limits but the result is incorrect:%
\begin{equation}
\int_{0}^{1}dx\int_{0}^{1}dy(L-x-y)=L-1\leq 0!  \tag{A7}
\end{equation}%
Two remarks are now in order. First, the presence of the theta function
after the $L$ itegration of PF of a HD system is general and not related to
the specific form of the $L$ integral. Second, if the integration over $L$
is extended to infinity, the theta function must still be present as the
integration from $0$ to the maximum condensate length (in the above example
it is 2) is included into it. In particular, the $\theta $ function will be
present if Laplace's transformation of the PF has been made before the
coordinate integration. In this case, to make the coordinate integrals all
having the same integration limits, one has no choice but to disregard the
theta function. However, as was shown above, omitting this function in the
result of Laplace's transformation is incorrect. Hence incorrect would be
the inverse Laplace's transformation, too.

Now we turn to the approach of Ref. \cite{Wojc}. To facilitate integration
over coordinates, the authors first Laplace transform the PF and miss the
theta function in the result. Due to this inconsistency, all the coordinate
integrals have independent integration limits and thus factorize. As a
consequence, some allowed points in the $N-1$ dimensional coordinate space
correspond to condensate's lengths exceeding the total length $L$ (e.g., in
our notations, all $\sigma $'s are equal to the disk diameter $d$ whereas $%
L<(N-1)d$). Next the inverse Laplace transformation is performed. However,
the total contribution to this integral comes from the single maximum point
which can lie within the domain restricted by the missing theta function.
Thus the intermediate inconsistency remains but, regarding for the
expressions for pressures which are similar to (\ref{PL}) and (\ref{PD}),
the result is correct. We emphasize that our approach is different from that
of Ref. \cite{Wojc}. We have not used Laplace transformation and our success
in performing the $\sigma $ (coordinate) integrals is due to the
representation (\ref{Teta}) of the theta function in terms of a delta
function.


\begin{thebibliography}{99}
\bibitem{Exchange} This exchanging event is also nicely illustrated
graphically in \cite{Robinson}.
\end{thebibliography}
\end{document}